\documentclass[aps,prl,twocolumn,superscriptaddress,noeprint]{revtex4-2}

\usepackage[english]{babel}
\usepackage{amsmath}
\usepackage{graphicx}
\usepackage{physics}
\usepackage{amssymb}
\usepackage[T1]{fontenc}
\usepackage[colorlinks=true, allcolors=blue]{hyperref}
\usepackage{indentfirst}
\usepackage{multirow}
\usepackage{kotex}
\usepackage{xcolor,colortbl}
\usepackage{makecell}
\usepackage{comment}
\usepackage{textcomp}
\usepackage{array}
\usepackage[utf8]{inputenc}
\usepackage{cleveref}
\crefname{equation}{Eq.}{Eqs.}
\Crefname{equation}{Equation}{Equations}
\crefname{figure}{Fig.}{Figs.}
\Crefname{figure}{Figure}{Figures}
\crefname{section}{Sec.}{Sects.}
\Crefname{section}{Section}{Sections}
\crefname{subsection}{subsection}{subsections}
\crefname{table}{Table}{Tables}
\crefname{appendix}{Appendix}{Apps.}
\Crefname{appendix}{Appendix}{Apps.}

\makeatletter
\def\maketitle{
	\@author@finish
	\title@column\titleblock@produce
	\suppressfloats[t]}
\makeatother 

\usepackage{natbib}

\newcolumntype{C}[1]{>{\centering\arraybackslash}p{#1}}
\renewcommand\thesection{\Roman{section}}

\begin{document}
\title{
\textbf{Hybrid qubit-oscillator module from motional states of two interacting atoms}
}

\author{Jaeyong Hwang}
\affiliation{JILA, NIST and Department of Physics, University of Colorado, Boulder, Colorado 80309, USA}
\affiliation{Center for Theory of Quantum Matter, University of Colorado, Boulder, Colorado 80309, USA}

\author{Tianrui Xu}
\affiliation{JILA, NIST and Department of Physics, University of Colorado, Boulder, Colorado 80309, USA}
\affiliation{Center for Theory of Quantum Matter, University of Colorado, Boulder, Colorado 80309, USA}
\affiliation{Institut Quantique and D\'epartement de Physique, Universit\'e de Sherbrooke, Sherbrooke, Qu\'ebec J1K 2R1, Canada}

\author{Sean R. Muleady}
\affiliation{Joint Center for Quantum Information and Computer Science,
University of Maryland and NIST, College Park, Maryland 20742, USA}
\affiliation{Joint Quantum Institute, University of Maryland and NIST, College Park, Maryland 20742, USA}

\author{Steven K. Pampel}
\affiliation{JILA, NIST and Department of Physics, University of Colorado, Boulder, Colorado 80309, USA}

\author{Gur Lubin}
\affiliation{JILA, NIST and Department of Physics, University of Colorado, Boulder, Colorado 80309, USA}

\author{Dawson P. Hewatt}
\affiliation{JILA, NIST and Department of Physics, University of Colorado, Boulder, Colorado 80309, USA}

\author{Cindy A. Regal}
\affiliation{JILA, NIST and Department of Physics, University of Colorado, Boulder, Colorado 80309, USA}

\author{Ana Maria Rey}
\affiliation{JILA, NIST and Department of Physics, University of Colorado, Boulder, Colorado 80309, USA}
\affiliation{Center for Theory of Quantum Matter, University of Colorado, Boulder, Colorado 80309, USA}

\date{\today}

\begin{abstract}

We propose a qubit-oscillator platform based on the motional states of two interacting atoms in an optical tweezer. By stroboscopically modulating an engineered trap with tunable anharmonicity, we implement a complete set of bosonic operations and their qubit-controlled counterparts with high fidelity. This motional control enables accurate detection of magnetic dipolar interactions with $\sim10$ Hz sensitivity in one second, reaching sub-Hz resolution within a few minutes in a $20\times20$ tweezer array under realistic experimental imperfections. Our approach establishes a versatile platform for motional quantum control of two atoms, with applications to spin-boson physics and precision sensing of interaction potentials and trapping environments.
\end{abstract}

\maketitle

Trapped neutral-atom systems, coupled via atom–atom interactions, have served as a foundational platform in atomic physics for several decades. Since the pioneering work on the manipulation of quantum degenerate gases~\cite{Bloch2008,Bloch2012,Schafer2020,Gross2021}, interest has steadily shifted toward achieving precise control over individual atoms, particularly using optical tweezers~\cite{Ashkin1997,Kaufman2021}. In these systems, strong and highly controllable Rydberg interactions between atoms in an array have become the leading way to couple qubits encoded in internal atomic levels~\cite{Saffman2010,Browaeys2020,Bluvstein2024}. Over the years, residual atomic motion in tweezers has remained as one of the sources of decoherence that limits the fidelity of neutral atom qubit gates, due to limitations in cooling and trapping ~\cite{Saffman2011,Tsai2025}. However, experiments are reaching a point at which control over motional bosonic modes is enabling new opportunities. For example, recent experimental efforts in optical tweezers have demonstrated full tomography of the motional degree of freedom of a single atom~\cite{Brown2023}, manipulation and squeezing of an atom's motional states~\cite{Brown2023,Lienhard2025}, generation of Laughlin states with two atoms~\cite{Lunt2024}, and the use of motional states for erasure detection of neutral atom qubits~\cite{Shaw2025}. These can be combined with theoretical efforts~\cite{Piotr2025,Huie2025} to harness the full potential of motional degrees of freedom (d.o.f.) in reconfigurable atom arrays. Despite these prospects, achieving full control over both internal and motional d.o.f. for a single neutral atom has remained challenging, partly due to competing spin-motion decoherence mechanisms, such as differential light shifts or sideband heating.

\begin{figure}[b]
 \centering
 \includegraphics[width=0.485\textwidth]{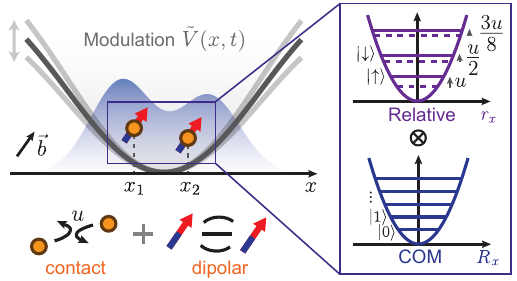}
 \caption{
 A hybrid qubit-oscillator module in the motion of two atoms trapped in a stroboscopically engineered optical trap. The qubit $\{\ket{\uparrow},\ket{\downarrow}\}$ resides in the two lowest motional states in the relative coordinate $r_x = (x_1-x_2)/\sqrt{2}$, energetically isolated by the nonlinearity generated via the contact interaction $\propto u$, while a harmonic oscillator mode $\{\ket{0},\ket{1},\cdots\}$ is encoded in the center-of-mass coordinate $R_x=(x_1+x_2)/\sqrt{2}$. Modulation of the potential $\tilde{V}(x,t)$ generates hybrid bosonic gates. As an example, the gate set can be used to perform precise measurements of on-site magnetic dipolar interactions, with the interaction strength varying with the magnetic field direction $\vec{b}$. 
 }
 \label{fig1}
\end{figure}

In this Letter, we propose a strategy to overcome this challenge by encoding a hybrid qubit-oscillator system entirely in the motional states of two neutral atoms, leveraging their intrinsic contact interaction. The oscillator is encoded in the center-of-mass (COM) mode, while the qubit is realized in the relative motional states through the nonlinearity arising from contact interaction. We demonstrate that a universal set of hybrid bosonic gates can be implemented by controlled modulation of an engineered optical tweezer potential. Beyond quantum control, this architecture enables interaction-resolved spectroscopy of weak dipolar couplings. Magnetic dipole-dipole interactions, for example, in ultracold atoms have traditionally been inferred through dipolar relaxation, spin dynamics, and their influence on many-body phases, particularly for strongly magnetic species~\cite{Pasquiou2011,dePaz2013,dePaz2014,Baier2016}, however, such approaches become increasingly challenging for weakly magnetic atoms. Magnetic-resonance techniques such as nuclear magnetic resonance (NMR) and electron spin resonance (ESR) provide an alternative route for interaction measurement via coherent phase accumulation~\cite{Abragam1961,Schweiger2001,Blanchard2015,Wilzewski2017}, but typically probe ensemble-averaged couplings in bulk systems. Here, the same hybrid-control toolbox used for quantum information processing encodes interaction-induced frequency shifts as phases in a relative-motion qubit and coherently transduces them into center-of-mass displacements for efficient readout. This approach combines the phase-sensitive character of magnetic-resonance spectroscopy with the microscopic control, deterministic geometry, and scalable programmability of optical tweezer arrays, enabling direct access to interactions between individual atom pairs while retaining the parallelism required for high-precision measurements. Because it relies on motional control and interaction-induced frequency shifts rather than species-specific spin dynamics, the protocol naturally extends from weakly magnetic alkali atoms to strongly dipolar atomic species.

\textit{Setup.---} We consider two bosonic atoms with mass $m$, confined in a single optical tweezer trap with high tunability of position and beam intensity. By rapidly flashing the beam between $j_{max}$ number of different positions with a frequency much faster than the tweezer trapping frequency, while applying appropriate trap depths in each position (see End Matter and Supplementary Material~\cite{Supp}), one can stroboscopically engineer~\cite{Alonso2013,Lacki2019,Yan2022} a time-averaged highly harmonic ``painted'' tweezer potential. The painted potential maintains strong transverse confinement along the $y$- and $z$-axes~\cite{Supp}, which freezes the atoms to their motional ground states, and restricts the dynamics to the $x$-axis.

To describe the motion of the two atoms in this potential, we define the COM coordinate as $R_x\equiv(x_1+x_2)/\sqrt{2}$ and the relative coordinate as $r_x\equiv (x_1-x_2)/\sqrt{2}$, where a factor $1/\sqrt{2}$ is used for convenience, and $x_i$ is the position of the $i$-th atom with $i=1,2$. In the position basis, the Hamiltonian for the harmonic trap is given by
\begin{equation} \label{eq:two_atom_Hamiltonian}
H = \sum_{q=R_x,r_x} \left[ -\frac{\hbar^2}{2m}\frac{\partial^2}{\partial q^2} + \frac{m\omega_x^2}{2}q^2 \right] + \sqrt{\pi}u \delta(r_x/x_0), \\
\end{equation}
where $u$ denotes the contact interaction strength of the ground motional state in the relative coordinate,  $x_0\equiv\sqrt{\hbar/m\omega_x}$ is the harmonic oscillator length, and $\omega_x$ is the net trap frequency equivalent to the trap frequency of a Gaussian potential with trap depth $V_0\equiv\frac{1}{4}m\omega_x^2W^2$ and $W$ the tweezer beam waist in the $x$-axis.

\textit{Qubit-oscillator system.---} 
We now express the dimensionless position operators in terms of bosonic annihilation operators $\hat{a}$ and $\hat{b}$ as $R_x/x_0\rightarrow \hat{R}_x\equiv (\hat{a}+\hat{a}^\dagger)/\sqrt{2}$ and $r_x/x_0\rightarrow \hat{r}_x\equiv (\hat{b}+\hat{b}^\dagger)/\sqrt{2}$, defining harmonic oscillator bases in the COM and relative coordinates. In the relative motional mode, due to the symmetry of indistinguishable bosons, only even excitation numbers are allowed by quantum statistics. Moreover, in the regime $u\ll\omega_x$, the interaction term $\sqrt{\pi}u\delta(\hat{r}_x)$ predominantly shifts the energy spectrum while leaving the motional wave functions highly unaffected. The resulting eigenenergies of the relative motional states, $\tilde{E}_{n=2l}$, are thus perturbed from $2l\hbar\omega_x$ with first-order corrections of $+u$, $+\frac{1}{2}u$, and $+\frac{3}{8}u$ for $l=0,1,2$, respectively, as shown in \cref{fig1}. The exact eigenvalues are analytically solvable~\cite{Busch1998,Idziaszek2005} and are used in our simulations. This allows us to isolate the two lowest relative motional states, $\{\ket{\uparrow},\ket{\downarrow}\}$, for qubit encoding. Therefore, \cref{eq:two_atom_Hamiltonian} reduces to a qubit-oscillator system of the form
\begin{equation} \label{eq_qubit_oscillator}
 \hat{H}_0 =~ \hbar\omega_x\hat{a}^\dagger\hat{a} -\frac{\hbar\tilde{\omega}}{2}\hat{\sigma}_z, \\
\end{equation}
where a qubit transition energy is $\hbar\tilde{\omega} \equiv \tilde{E}_{2} - \tilde{E}_0 \simeq 2\hbar\omega_x-u/2$, and $\hat{\sigma}_z\equiv\ket{\uparrow}\bra{\uparrow}-\ket{\downarrow}\bra{\downarrow}$ (See Fig.~\ref{fig1}). We note that the anharmonicity suppressing leakage out of the qubit subspace, given by $|\tilde{E}_4-2\tilde{E}_2+\tilde{E}_0|$, reaches a maximum value $0.084\hbar\omega_x$ at $u/\hbar\omega_x=0.61$, across all trap depths.

\begin{figure}[b]
 \centering\includegraphics[width=0.49\textwidth]{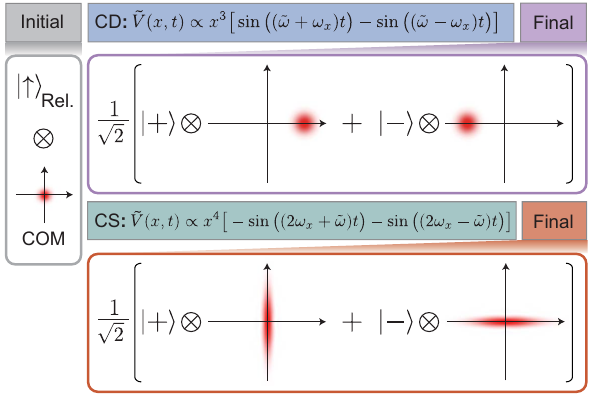}
 \caption{
 Implementation of controlled displacement (CD) and controlled squeezing (CS) gates from the initial state $\ket{\uparrow}\ket{0}$. Multi-frequency driving of nonlinear terms enables COM–relative coupling gates.  
 }
 \label{fig2}
\end{figure}

\begin{figure*}[t]
 \centering
 \begin{minipage}[b]{1.0\linewidth} 
 \centering
\renewcommand{\arraystretch}{1.1}
\setlength{\extrarowheight}{3pt}
  \begin{tabular}[b]{|m{0.05\textwidth}<{\centering}||m{0.15\textwidth}<{\centering}|m{0.15\textwidth}<{\centering}|}
 \hline
 \multirow{3}{*}{\makecell{Gate \\ \\ ~}} & \multirow{2}{*}{\makecell{Gate Generator: \\ $-i\hat{H}_It/\hbar$ }} & \multirow{2}{*}{\makecell{Infidelity: $1-\mathcal{F}$\\(examples)}} \\
  & & \\
 \hline
 D & $\alpha\hat{a}^\dagger-\alpha^*\hat{a}$  & $\lesssim10^{-5}~(|\alpha|=3)$ \\ 
 \hline
 R & $i\gamma\hat{a}^\dagger\hat{a}+i\gamma'\hat{\sigma}_z$ & $\lesssim10^{-4}~(\gamma=\frac{\pi}{2})$\\ 
 \hline
 QR & $-i\gamma\hat{\sigma}_\phi/2$  & $\lesssim10^{-3}~(\gamma=\frac{\pi}{2})$\\ 
 \hline
 S & $\frac{1}{2}(\xi^*\hat{a}^2-\xi\hat{a}^{\dagger2})$  & $\lesssim10^{-5}~(|\xi|=1)$ \\ 
 \hline
 CD & $(\alpha\hat{a}^\dagger-\alpha^*\hat{a})\hat{\sigma}_\phi$ & $\lesssim10^{-0.8}~(|\alpha|=3)$ \\ 
 \hline
 CR & $(i\gamma \hat{a}^\dagger\hat{a}) \hat{\sigma}_\phi$ & $\lesssim10^{-2}~(\gamma=\frac{\pi}{2})$\\ 
 \hline
 CS & $\frac{1}{2}(\xi^*\hat{a}^2-\xi\hat{a}^{\dagger2})\hat{\sigma}_\phi$ & $\lesssim10^{-4}~(|\xi|=1)$\\ 
 \hline
 \end{tabular}
\renewcommand{\arraystretch}{1}
 \centering
 \includegraphics[width=0.51\textwidth]{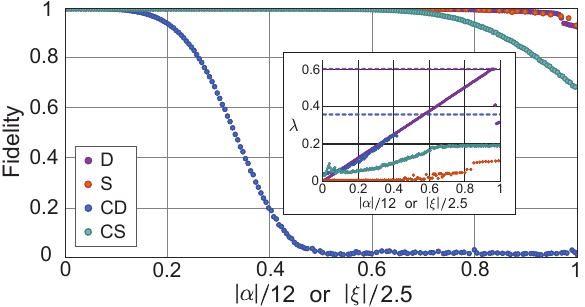}
 \end{minipage}
 \caption{
 A list of native gates: displacement (D), rotation (R), qubit rotation (QR), squeezing (S), controlled displacement (CD), controlled rotation (CR), and controlled squeezing (CS). Left: Gate infidelities for \{D, S, CD, CS\} evaluated from the initial ground state $\ket{\uparrow}\ket{0}$ using representative parameters ($|\alpha|=3$, $|\xi|=1$), and for \{R, QR, CR\} from the displaced state $\ket{\uparrow}\ket{\alpha{=}1}$ with rotation angle $\gamma=\pi/2$; see~\cite{Supp} for details. Right: Numerical gate fidelities for \{D, S, CD, CS\}. Gates \{D, CD\} use $j_{\max}=5$, while \{S, CS\} use $j_{\max}=3$. Each point corresponds to the maximal fidelity optimized over modulation strengths $\lambda$. Inset: optimum values of $\lambda$ used. The dotted lines mark the upper limits for \{D, CD\}, beyond which a trap depth becomes negative and thus nonphysical.
 }
\label{fig3}
\end{figure*}

\textit{Quantum gate generation.---} In addition to the base Hamiltonian $\hat{H}_0$, the potential can be further engineered to include a time-varying part.
By adjusting the temporal modulation of the trap depths at each position, one can engineer a tunable one-dimensional (1D) effective potential of the form (See End Matter for details):
\begin{equation} \label{eq:potential_general}
V(x_i,t) \simeq \sum_{k=1}^{k_{max}} V_k(t)(x_i/W)^k. \\
\end{equation}
The controllability, indicated by $k_{max}$, generally increases with increasing $j_{max}$. In this work, we choose a minimal setting with $k_{max}=5$, which can be achieved with $j_{max}=3$ or $5$~\cite{Supp}. 
It is convenient defining the vector $\{V_k(t)\}=\{0,2V_0,0,0,0\}+\lambda\{\tilde{V}_k(t)\}$ characterizing the full potential, where $\tilde{V}_k(t)$ are slow modulations used to implement specific gate operations, and the tunable parameter $\lambda$ sets their overall strengths. 

The full Hamiltonian is $\hat{H} = \hat{H}_0 + \lambda \sum_{i=1,2}\tilde{V}(x_i,t)$ with the modulation $\tilde{V}(x_i,t)=\sum_{k=1}^{k_{max}}\epsilon_x^k \tilde{V}_k(t) \hat{x}_i^k$ and $\epsilon_x\equiv x_0/W$. Going to the interaction picture set by $\hat{H}_0$, we choose the modulation frequencies such that only the desired terms become stationary, while all undesired terms oscillate rapidly in time, thus rotated out. For example, with $\{\tilde{V}_k(t)\}=\{V_0\sin(\omega_xt-\theta),0,0,0,0\}$, the only term that survive in the interaction Hamiltonian $\hat{H}_I$ generates a bosonic displacement gate $\hat{D}(\alpha)\equiv \exp\left( \alpha\hat{a}^\dagger - \alpha^*\hat{a} \right)$, where $\alpha\propto \lambda\epsilon_xV_0Te^{i\theta}$ with $\theta$ setting the direction of the phase-space displacement.

Furthermore, simultaneously modulating higher-order terms with two different frequencies, such as $\{\tilde{V}_k(t)\} = \{0,0,\tilde{V}_3(t),0,0\}$ with $\tilde{V}_{3}(t)=\sum_{\eta=\pm}\eta V_0\sin\left((\tilde{\omega}+\eta\omega_x)t-\eta\theta-\phi\right) $, couples the COM and the relative modes (see \cref{fig2}). This specific case generates an effective unitary operation of the form 
\begin{align}
 \exp(-i\hat{H}_IT/\hbar) &= \exp((\alpha\hat{a}^\dagger-\alpha^*\hat{a})\otimes\hat{\sigma}_\phi),
\end{align} which corresponds to a controlled displacement (CD) operation with $\alpha\propto \lambda\epsilon_x^3V_0Te^{i\theta}$ after operation time $T$. Here, $\hat{\sigma}_\phi\equiv\hat{\sigma}_x\cos\phi+\hat{\sigma}_y\sin\phi$ acts on the control qubit with angle $\phi$ that sets the rotation axis in the $x$-$y$ plane.
For example, a CD gate with $\phi=0$ transforms an initial state $\ket{\uparrow}\ket{0}$ into $\frac{1}{\sqrt{2}}\left(\ket{+}\ket{\alpha}+\ket{-}\ket{-\alpha}\right)$, where $\ket{\pm}\equiv(\ket{\uparrow}\pm\ket{\downarrow})/\sqrt{2}$ and $\ket{\alpha}\equiv\hat D(\alpha)\ket{0}$, or equivalently $\frac{1}{2}\ket{\uparrow}(\ket{\alpha}+\ket{-\alpha})+\frac{1}{2}\ket{\downarrow}(\ket{\alpha}-\ket{-\alpha})$. 
Similarly, a controlled squeezing operation $\mathrm{CS}=\exp[\tfrac{1}{2}(\xi^*\hat{a}^2-\xi\hat{a}^{\dagger 2})\hat{\sigma}_x]$, producing superpositions of squeezed states, can be realized by modulating the $x^4$ term with frequencies $2\omega_x\pm\tilde{\omega}$. 
All possible native quantum gates are shown in the left panel of \cref{fig3} and the supplementary material.

We evaluate the performance of all the proposed quantum gates via full evolution of the time-\emph{dependent} Schr\"odinger equation. To analyze gate performance, we compute the gate fidelity, defined as 
\begin{align}
\begin{split}
 \mathcal{F} &= \left|\bra{\psi_0}\hat{\mathcal {U}}^\dagger\mathcal{T}\left(e^{-\frac{i}{\hbar}\int_0^T\hat{H}_I(t)dt}\right)\ket{\psi_0}\right|^2, \\
\end{split}
\end{align}
where $\mathcal{T}$ is the time-ordering operator, the initial state is $\ket{\psi_0}=\ket{\uparrow}\ket{0}$, and $\hat{\mathcal{U}}$ is the target unitary operator of the desired gate operation. The interaction Hamiltonian $\hat{H}_I(t)$ is the \emph{full} 3D time-dependent Hamiltonian, including \emph{all} time-dependent terms and \emph{all} higher-order $\mathcal{O}((x_i/W)^{k_{max}+1})$ terms hidden in \cref{eq:potential_general}; see End Matter for the specific parameters used in our simulations. In \cref{fig3}, we report the optimum fidelity of each gate as a function of the displacement (squeezing, rotation) amplitudes $|\alpha|$ ($|\xi|$, $|\gamma|$), obtained by tuning the modulation strength $\lambda$ for each amplitude to minimize both higher-order contributions $V_0\sum_i\mathcal{O}((x_i/W)^{k_{max}+1})$ and leakage outside the qubit subspace. 
Despite our minimal setting with $j_{\max}\leq 5$, we observe that all native gates in Fig.~\ref{fig3} achieve fidelities above $99\%$ over a wide parameter range in the noiseless case. The CD gate retains high fidelity only for moderate displacements, $|\alpha|\lesssim2$, though this limit could be substantially extended using improved potential-painting techniques with larger $j_{max}$ demonstrated in previous works~\cite{Alonso2013,Lacki2019,Yan2022}.
The timescales for optimal performance across all gates and all gate parameters $\{\alpha,\xi,\gamma\}$ are roughly 1-10 $\mu$s for the D gate, 0.1-1 ms for the $\{$S,QR,R$\}$ gates, 0.1-10 ms for the CD gate, and 1-10 ms for the $\{$CR,CS$\}$ gates, in sub-mK to mK trap depths. Note that different choices of $\lambda$, especially ones larger than optimal choices, may slightly reduce the fidelity but can increase the gate speed, which may be preferable experimentally.

\textit{Sensing Application.---} As an explicit application, we now show that the motional control of two atoms enables the accurate detection of weak magnetic dipole-dipole interaction. Two magnetic dipoles with the 3D relative position vector $\vec{r}$ and distance $r=|\vec{r}|$, interact via the following dipolar potential: 
\begin{equation} \label{eq:dipolar_interaction}
V_{d}(\vec{r}) = \frac{\mu_0 \mu_m^2}{4\pi r^3} \left(1 - 3(\vec{r}\cdot \vec{b})^2/r^2\right),
\end{equation}
where $\mu_m=\mu_Bg_F$ is the magnetic moment and $g_F$ the Land\'e g-factor. For example, for  $^{87}$Rb atoms in the hyperfine state $\ket{5^2S_{1/2},F=1,m_F=1}$, $g_F$ is $-1/2$. Here $\vec{b}$ is a unit vector along the magnetic dipole moment set by the direction of an external magnetic field. 
We are interested in the energy difference, $\hbar \nu\equiv(V_{d,\uparrow} - V_{d,\downarrow})$, between the dipolar interaction energies $V_{d,\uparrow}$ and $V_{d,\downarrow}$ experienced by the qubit states, $\ket{\Psi_\uparrow}\equiv\ket{\uparrow}\ket{\psi_{0,y}}\ket{\psi_{0,z}}$ and $\ket{\Psi_\downarrow}\equiv\ket{\downarrow}\ket{\psi_{0,y}}\ket{\psi_{0,z}}$, where $\ket{\psi_{0,y}}$ ($\ket{\psi_{0,z}}$) is the motional ground state on the axis $y$- ($z$-) of our 3D trap. For our  painted potential, $\nu$ can vary from zero to $\sim 2\pi\times119~\text{Hz}\times(V_0~/~1~\text{mK})^{3/4}$, depending on the  $\vec{b}$ orientation~\cite{Supp}.

\begin{figure}[b]
 \centering
 \includegraphics[width=0.49\textwidth]{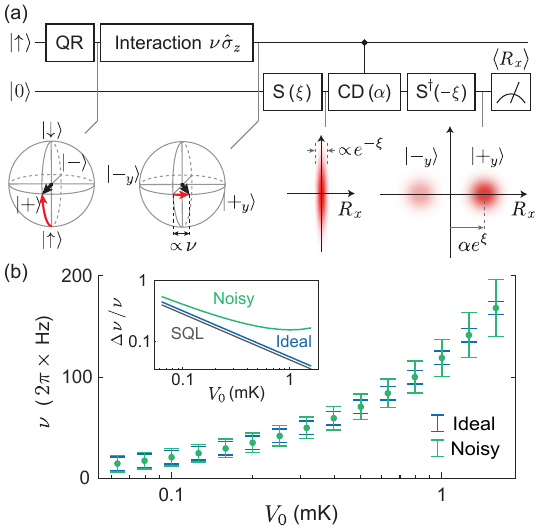}
 \caption{
    Magnetic dipole-dipole interaction sensing with motional states. (a) Sensing protocol based on hybrid bosonic gates and COM quadrature measurement. The difference in dipolar interaction strength, $\nu$, between $\ket{\uparrow}$ and $\ket{\downarrow}$, is detected by mapping the interaction-induced phase onto the COM motion using CD gate enhanced by squeezing and unsqueezing operations.
    (b) Sensitivities (error bars) in the ideal case reaching the standard quantum limit (SQL), and in the presence of uncertainties in the level structure or system parameters.
 }
 \label{fig4}
\end{figure}
We now show that hybrid bosonic gates can significantly simplify interaction sensing by straightforward time-of-flight (TOF) measurements without the need for full state tomography. As illustrated in \cref{fig4}, the protocol begins with the two atoms prepared in the motional ground state of the tweezer, corresponding to the relative-motion qubit state $\ket{\uparrow}$ and the COM oscillator ground state $\ket{0}$. After preparing the superposed state $\ket{+}\ket{0}$ with applying qubit-rotation gate $\hat{U}_{\rm QR}=\exp(i\pi\hat{\sigma}_y/4)$, 
the system is subsequently allowed to evolve freely for a dark time $t_d$. During this period, the magnetic dipolar interaction induces a small shift $\nu$ of the qubit transition frequency, leading to a phase accumulation corresponding to a rotation by an angle $\nu t_d$ about the $\hat{\sigma}_z$ axis of the Bloch sphere. Consequently, the interaction strength is encoded in the population imbalance between the states $\ket{\pm_y}$, where $\ket{\pm_y}\equiv (\ket{\uparrow}\pm i\ket{\downarrow})/\sqrt{2}$.

To read out this small phase, we coherently transfer the population information in the relative-motion qubit to the center-of-mass (COM) oscillator using the controlled-displacement gate $\hat{U}_{\rm CD}=\exp\left[\alpha\hat{\sigma}_y(\hat{a}^\dagger-\hat{a})\right]$. A key capability of our hybrid-control toolbox is the preparation of nonclassical oscillator states. By placing the controlled-displacement gate between two squeezing operations with amplitudes $\xi$ and $-\xi$~\cite{Burd2019}, the qubit-dependent displacement is amplified according to $\pm\alpha\rightarrow\pm\alpha e^{\xi}$. The squeezing therefore acts as a metrological resource that enhances the transduction of the interaction-induced phase into a measurable motional signal. As a result, the phase accumulated during the dark evolution is converted into an average COM displacement $\langle R_x\rangle \propto \alpha e^{\xi}\sin(\nu t_d)$, which can be directly measured by time-of-flight imaging with high precision. 
The variance of measured $\nu$ is thus~\cite{Supp}
\begin{equation}
\Delta^2\nu = \frac{1}{N_t} \frac{\langle R_x^2 \rangle - \langle R_x\rangle^2}{\left| \partial \langle R_x\rangle /\partial \nu \right|^2}= \frac{1}{N_t t_d^2} \left( 1 + \frac{1}{4\alpha^2e^{2\xi}} \right),
\end{equation}
where $N_t=MN$ is the total number of repeated measurements, with $M$ the number of measurements and $N$ the number of identical copies in each measurement. 
The sensitivity $\Delta \nu$ is shown by the error bars in \cref{fig4}(b) for both the ideal case (without experimental imperfections) and the noisy case that includes the dominant noise source in our protocol: stochastic shot-to-shot tweezer-parameter fluctuations, modeled as 0.2\% intensity noise in a single trap, consistent with state-of-the-art homogeneity of tweezer arrays~\cite{Spar2022,Lis2023,Chew2024,Ammenwerth2025}.
In the simulations shown in \cref{fig4}, we assume an $N=20\times20$ array, with $M=5$ repetitions achievable within 1 second. Even in the presence of realistic noise and experimental imperfections, the protocol retains sensitivity to interaction shifts at the $\sim$10 Hz level and is projected to achieve sub-Hz resolution within a few minutes of averaging.

\textit{State preparation and readout.---} To prepare the initial state $\ket{\uparrow}\ket{0}$, one potential laser-cooling path is to adiabatically merge two cold atoms, initially confined in separate optical tweezers, into a single trap with negligible heating~\cite{Pampel2025}, followed by additional Raman sideband cooling while avoiding photoassociation processes~\cite{Urvoy2019}.
The average COM displacement can be measured via standard TOF imaging, while full quantum tomography of the motional state of a single atom~\cite{Home2020} can be achieved by applying $(\hat{a}+\hat{a}^\dagger)\hat{S}_x$ that couples the motional modes to an additional atomic internal state via Raman transitions. Here $\hat{S}_{x,y,z}$ are spin operators acting on the \emph{internal} spin, followed by measuring $\langle\hat{S}_z\rangle$. Similarly, in the case of two neutral atoms, relevant information about the quantum state can be extracted through evolution under $(\hat{a}_1+\hat{a}^\dagger_1)\hat{S}_{x,1}+(\hat{a}_2+\hat{a}^\dagger_2)\hat{S}_{x,2}$, followed by population measurements such as $\langle\hat{S}_{z,1}\hat{S}_{z,2}\rangle$~\cite{Supp}.

\textit{Conclusion.---}
We have proposed protocols that leverage native contact interactions between two atoms confined in a stroboscopically engineered optical tweezer to realize a hybrid qubit-oscillator module encoded entirely in motional degrees of freedom. Using this platform, we demonstrate a universal set of hybrid bosonic operations and show how they enable interaction-resolved spectroscopy of weak magnetic dipolar interactions. Extensions to multiple motional modes and richer relative-coordinate manifolds may further enhance sensitivity to anisotropic interactions and enable applications in quantum simulation of molecular systems~\cite{Kang2024,Navickas2025,So2025}, while nonclassical bosonic encodings such as cat states~\cite{Grimm2020,Puri2020} could improve robustness against imperfections and decoherence. More generally, the framework can be extended to few-body systems and many-particle encodings, potentially including topologically protected architectures~\cite{Lunt2024}.

\textit{Acknowledgments.---}We acknowledge useful discussions and feedback on the manuscript from Conall McCabe, Joanna Lis, and Adam Kaufman. We thank Youcef Baamara and Piotr Grochowski for fruitful discussions on the project. This work is supported by the NSF JILA-PFC PHY-2317149 and NSF QLCI award OMA-2016244, the U.S. Department of Energy, Office of Science, National Quantum Information Science Research Centers, Quantum Systems Accelerator, and NIST. S.R.M. is supported by the NSF QLCI award OMA-2120757.

\bibliography{references}

\onecolumngrid
\vspace*{5px}
\begin{center}\textbf{  End Matter} \end{center}
\vspace*{5px}
\twocolumngrid

\section{Potential painting:
Engineering an effective 1D potential from a 3D tweezer}

\textit{Setup.---}We consider two bosonic atoms, with s-wave scattering length $a_s$ and mass $m$, trapped in a three-dimensional (3D), time-dependent potential $V_{3D}(\vec{x}_i,t)$. We label the position of each atom as $\vec{x}_i=\{x_i,y_i,z_i\}~(i=1,2)$, and define the COM coordinate as $\vec{R}=\frac{1}{\sqrt{2}}(\vec{x}_1+\vec{x}_2)=\{R_x,R_y,R_z\}$ and the relative coordinate as $\vec{r}=\frac{1}{\sqrt{2}}(\vec{x}_1-\vec{x}_2)=\{r_x,r_y,r_z\}$. In the position basis, the two-atom Hamiltonian is given by
\begin{align}
\begin{split} \label{eq1}
 {H} &= \sum_{i=1,2} \left[ -\frac{\hbar^2}{2m}\nabla^2_i + V_{3D}(\vec{x}_i,t) \right] + g\delta^{(3)}(\vec{r})\frac{\partial}{\partial |\vec{r}|}|\vec{r}| ,\\
\end{split}
\end{align}
where $g=\sqrt{2}\pi\hbar^2a_s/m$, and the $\frac{\partial}{\partial |\vec{r}|}|\vec{r}|$ term is introduced for mathematical consistency, handling the divergence of the molecular bound state~\cite{Busch1998}; however, when acting
on wavefunctions that are regular at the origin as the ones considered here, the regularization operator  has no effect 
and can be dropped. 

To manipulate the motional states of the two trapped atoms, we propose using a stroboscopically engineered ``painted'' optical potential~\cite{Alonso2013,Lacki2019,Yan2022}. For simplicity, we consider rapidly flashing a single beam between $j_{max}$ different positions, $x/W\in\{\zeta_1,\zeta_2,\cdots,\zeta_{j_{max}}\}$, with trap depth $j_{max}U_j(t)$ at position $\zeta_j$, where $W$ is the beam waist. The potential stays at each position for the same amount of time, $T/j_{max}$ with $T\ll2\pi/\omega_x$, so that the effective potential is time-averaged over all $j$'s with each trap depth normalized by value $j_{max}$. At the same time, the time dependence of $U_j(t)$ is at a similar timescale set by the trap frequency and thus cannot be time-averaged. This yields a time-effective trapping potential
\begin{equation}
\label{eq_Gaussian_beam}
 V_{3D}(\vec{x}_i,t) = \sum_{j=1}^{j_{max}} \frac{-U_j(t)}{1+z_i'^2}\exp\left[ \frac{-2(x_i'-\zeta_j)^2-2y_i'^2}{1+z_i'^2} \right] 
\end{equation}
with dimensionless parameters defined as $x'_i\equiv x_i/W$, $y'_i\equiv y_i/W$, and $z'_i\equiv z_i/z_R$ for Rayleigh range $z_R$. 

We work in a regime where the time-effective trapping potential is anisotropic, i.e. $\omega_y,\omega_z>\omega_x$, where $\omega_\mu$ is the effective trap frequency along the $\mu\in \{x,y,z\}$-axis, with corresponding oscillator length $\mu_0 \equiv \sqrt{\hbar/m\omega_\mu}$. The usual Gaussian potential has the relation $\omega_z<\omega_x\sim\omega_y$, but stroboscopically distributing trap centers along $x$-direction can result in a smaller effective $\omega_x$ compared to the natural trap frequency, making it feasible to energetically isolate the $x$-direction. Furthermore, one could also consider adding additional trapping beams to induce stronger confinement in the $y$- and $z$-directions to improve control along these dimensions by keeping $\omega_y$ and $\omega_z$ large and constant in time. In both cases, the 3D potential $V_{3D}(\vec{x_i},t)$ effectively becomes 1D:
\begin{equation} \label{eq_sum_of_traps}
 V(x_i,t) = \int dy_idz_i |\psi_{0,y}(y_i)|^2|\psi_{0,z}(z_i)|^2V_{3D}(\vec{x}_i,t),\\
\end{equation} 
where the state in $y$- ($z$-) direction is approximated to be $\psi_{0,y}(y_i)$ ($\psi_{0,z}(z_i)$), the ground state wavefunction of the $i$-th atom in a harmonic potential with frequency $\omega_{y}$ ($\omega_{z}$).

\section{Simulation parameters}

In our simulations, we use parameters relevant for $^{87}$Rb atoms with a tweezer beam waist $W=700$ nm, trap frequency $\omega_x=2\pi\times140$ kHz ($V_0=1$ mK), and a perturbation parameter $\epsilon_x=0.041$. To reduce qubit leakage by increasing anharmonicity, we set the interaction strength to $u/\hbar\omega_x=0.86$ for $[\rm D,CD]$ gates and $u/\hbar\omega_x=0.36$ for $[\rm S,CS]$ gates by adjusting $\omega_y$ and $\omega_z$, both of which are experimentally relevant. We obtain optimal combinations of trap positions and trap depths for each gate, and report the gate fidelities $\mathcal{F}$ with optimal choices of modulating strength $\lambda$. 

For the simulation parameters of the results in Fig.~\ref{fig3}, for the D and CD gates, we set $j_{max} = 5$ with beam center positions $\{\zeta_j\}=\{-1.14,-0.56,0.04,0.38,0.90\}$, and trap depths ${\bf U}^{(0)}/V_0=\{1.81,2.94,2.82,1.42,2.88\}$ to generate the initial highly harmonic potential. The D gate is generated by applying trap depths ${\bf U}(t)/V_0={\bf U}^{(0)}/V_0+\{3.01,2.00,2.04,0.49,2.16\}\times\lambda \sin(\omega_xt)$, while the CD gate is obtained by applying trap depths ${\bf U}(t)/V_0={\bf U}^{(0)}/V_0+\{5.04,3.16,3.83,1.21,4.91\}\times\lambda [\sin((\tilde{\omega}+\omega_x)t) - \sin((\tilde{\omega}-\omega_x)t)]$. Similarly, for the S and CS gates, we set $j_{max}=3$ with beam center positions $\{\zeta_j\}=\{-0.6775,0,0.6775\}$, and trap depths ${\bf U}^{(0)}/V_0=\{1.76,2.17,1.76\}$ for the initial harmonic potential. The S gate is then implemented by applying ${\bf U}(t)/V_0={\bf U}^{(0)}/V_0-\{0.88,1.09,0.88\}\times\lambda\sin(2\omega_xt)$, while the CS gate uses ${\bf U}(t)/V_0={\bf U}^{(0)}/V_0-\{0.88,0.59,0.88\}\times\lambda [\sin((2\omega_x+\tilde{\omega})t) - \sin((2\omega_x-\tilde{\omega})t)]$. 

As an example, for $|\alpha|=3$ or $|\xi|=1$, with optimized $\lambda$'s we obtain gate infidelities ($1-\mathcal{F}$) for the $[\rm D,S,CD,CS]$ gates of $[6.2\times10^{-7},1.3\times10^{-5},1.7\times10^{-1},1.1\times10^{-4}]$, as summarized in Fig.~\ref{fig3}, with corresponding gate times $[7.1,190,5100,13000]~\mu s$.

\setcounter{equation}{0}
\setcounter{figure}{0}
\setcounter{table}{0}
\setcounter{section}{0}
\makeatletter

\renewcommand{\theequation}{S\arabic{equation}}
\renewcommand{\thefigure}{S\arabic{figure}}
\renewcommand{\thetable}{S\arabic{table}}
\renewcommand{\thesection}{S\arabic{section}}

\title{
\textbf{Supplemental Material for ``Hybrid qubit-oscillator module from motional states of two interacting atoms''}
}

\clearpage
\maketitle
\onecolumngrid

\section*{
Engineering an effective 1D potential from a 3D optical tweezer } 

The 3D Gaussian potential at the beam center $\{x,y,z\}=\{\zeta_jW,0,0\}$ with trap depth $j_{max}U_j(t)$ can be modelled as
\begin{align}
\begin{split}
 \hat{f}(x,y,z;U_j(t),\zeta_j) &= -\frac{j_{max}U_j(t)}{1+\epsilon_z^2\hat{z}^2}\exp\left[\frac{-2(\epsilon_x\hat{x}-\zeta_j)^2-2\epsilon_y^2\hat{y}^2}{1+\epsilon_z^2\hat{z}^2} \right] ,\\
\end{split}
\end{align}
with small parameters $\epsilon_x\equiv x_0/W$, $\epsilon_y\equiv y_0/W$, and $\epsilon_z\equiv z_0/z_R$. The operators $\hat{x}$, $\hat{y}$, $\hat{z}$ represent the dimensionless positions $x'$, $y'$, $z'$ in the main text, and $\zeta_j$ is dimensionless position defined in the End Matter. The optical tweezer is rapidly flashed among various positions $\{\zeta_j\}$ for $j=1,\cdots,j_{max}$, so that the overall 3D potential in the slow time scale (in the time scale of trap frequency) becomes
\begin{align}
\begin{split}
 V_{3D}(x,y,z) = \sum_{j=1}^{j_{max}} \frac{1}{j_{max}}\hat{f}(x,y,z;U_j(t),\zeta_j). \\
\end{split}
\end{align}
Assuming an atom confined to the ground states in $y$- and $z$-directions with trap frequencies $\omega_y$ and $\omega_z$, we get
\begin{align}
\begin{split}
  V_{1D}(x)=\bra{0}_y\bra{0}_zV_{3D}(x,y,z)\ket{0}_z\ket{0}_y &= U_j(t) \sum_{m=0}^{\infty} c_m(\zeta_j,\epsilon_y,\epsilon_z)\epsilon_x^m\hat{x}^m, \\
\end{split}
\end{align}
and 
\begin{align}
\begin{split}
 c_m(\zeta_j,\epsilon_y,\epsilon_z) &\equiv (-1)^{m+1}\sum_{s=\left\lfloor\frac{m+1}{2}\right\rfloor}^{\infty}\left(\sum_{t=0}^\infty (2\epsilon_y^2)^t{-1/2 \choose t}g(s+t+1,\epsilon_z)\right) {2s \choose m}\frac{(-2)^s}{s!}\zeta_j^{2s-m}, \\
\end{split}
\end{align}
where ${-1/2 \choose t}=\frac{(2t-1)!!}{(-2)^tt!}$, $(-1)!!=1$, and $g(n,a)\equiv\pi^{-1/2}\int_{-\infty}^\infty (1+a^2x^2)^{-n}e^{-x^2}dx$. When $a\ll 1$, $g(n,a)\approx 1$ for small $n$. The above expression thus corresponds to the 1D potential, where its shape can be engineered by choosing appropriate $\{\zeta_j\}$ and $\{U_j(t)\}$.

\section*{Trap Depth Design }

In general, up to $k_{max}=j_{max}$ number of $V_k(t)$ terms can be controlled by $j_{max}$ number of non-negative trap depths $U_j(t)$, determined by ${\bf U}(t)={\bf C}^{-1}{\bf V}(t)$. However, in the case of a spatially symmetric effective potential, a larger $k_{max}$ can be achieved with the same $j_{max}$. In this case, the values of $\zeta_j$ are symmetrically distributed to the left and the right of their average value, namely, $U_j(t)=U_{j_{max}+1-j}(t)$ and $\zeta_j=-\zeta_{j_{max}+1-j}$ for all $j$'s, resulting in $V_k(t)=0$ for all odd $k$'s, and thus up to $k_{max}=2\lceil j_{max}/2\rceil+1$ is deterministically set by ${\bf U}(t)$. 

Below, we show how to set amplitudes ${\bf V}(t)={\bf V}_0+\lambda\tilde{{\bf V}}(t)=\{0,2V_0,0,0,0\}+\lambda\{\tilde{V}_1(t),\tilde{V}_2(t),\tilde{V}_3(t),\tilde{V}_4(t),\tilde{V}_5(t)\}$ for $k_{max}=5$ in our work, which corresponds to $j_{max}=3$ for the symmetric case and $j_{max}=5$ for the general case. 

\subsection{i) Symmetric potential ($j_{max}=3$ for S, R, QR, CS, CR gates)}

The bosonic squeezing and rotation gates are engineered with three flashing positions with $\zeta_3=-\zeta_1=\zeta$, $\zeta_2=0$, and trap depth at each position as
\begin{align}
\begin{split}
 U_2(t) &= \frac{c_4(\zeta)\big(2V_0+\lambda\tilde{V}_2(t)\big)-c_2(\zeta)\lambda\tilde{V}_4(t)}{c_4(\zeta)c_2(0)-c_4(0)c_2(\zeta)}, \\
 U_1(t) = U_3(t) &= \frac{1}{2}\times\frac{-c_4(0)\big(2V_0+\lambda\tilde{V}_2(t)\big)+c_2(0)\lambda\tilde{V}_4(t)}{c_4(\zeta)c_2(0)-c_4(0)c_2(\zeta)},
\end{split}
\end{align} 
where $c_m(\zeta)$ is a shortened term for $c_m(\zeta,\epsilon_y,\epsilon_z)$. Additionally, it is possible to find the optimal $\zeta$ value to minimize the 6th-order term $\hat{x}^6$. We use $\zeta=0.6775$ in our simulations, assuming $W=700$nm, $V_0/k_B=1$mK, $[\omega_x,\omega_y,\omega_z]=2\pi\times[140,709,195]$kHz, and $[\epsilon_x,\epsilon_y,\epsilon_z]=[0.041,0.018,0.014]$ for trapped $^{87}$Rb atoms. 

\subsection{ii) General potential ($j_{max}=5$ for D, CD gates)}

The bosonic displacement gates are engineered with five flashing positions that generate time-modulation terms ${\bf \tilde{V}}(t)=\{\tilde{V}_1(t),0,\tilde{V}_3(t),0,0\}$. With the dimensionless center positions $\{\zeta_j\}$ and trap depths $\{U_j(t)\}$ ($j=1,\cdots,5$), we obtain $U_j(t)$ via
\begin{align}
\begin{split}
 \begin{pmatrix} U_1(t) \\ U_2(t) \\ U_3(t) \\ U_4(t) \\ U_5(t) \end{pmatrix} &= \begin{pmatrix} c_1(\zeta_1) & c_1(\zeta_2) & c_1(\zeta_3) & c_1(\zeta_4) & c_1(\zeta_5) \\ c_2(\zeta_1) & c_2(\zeta_2) & c_2(\zeta_3) & c_2(\zeta_4) & c_2(\zeta_5) \\ c_3(\zeta_1) & c_3(\zeta_2) & c_3(\zeta_3) & c_3(\zeta_4) & c_3(\zeta_5) \\ c_4(\zeta_1) & c_4(\zeta_2) & c_4(\zeta_3) & c_4(\zeta_4) & c_4(\zeta_5) \\ c_5(\zeta_1) & c_5(\zeta_2) & c_5(\zeta_3) & c_5(\zeta_4) & c_5(\zeta_5) \end{pmatrix}^{-1} \begin{pmatrix} \lambda\tilde{V}_1(t) \\ 2V_0 \\ \lambda\tilde{V}_3(t) \\ 0 \\ 0 \end{pmatrix}, \\
\end{split}
\end{align} 
where we use $\{\zeta_1,\zeta_2,\zeta_3,\zeta_4,\zeta_5\}=\{-1.14,-0.56,0.04,0.38,0.90\}$ for our simulations, and $c_m(\zeta_j)$ is again $c_m(\zeta_j,\epsilon_y,\epsilon_z)$ defined above.

\section*{Two interacting atoms in a harmonic trap}

The Hamiltonian describing two interacting atoms in a 1D harmonic trap is given by Eq.1 in the main text. Since the center of mass and relative coordinates decouple, we focus here on the relative coordinates. The Hamiltonian in the relative coordinate along the $x$-axis can thus be written as
\begin{align}
\begin{split}
 \hat{H} = \hbar\omega_x \hat{b}^\dagger\hat{b} + \sqrt{\pi}u\delta(\hat{r}_x), \\
\end{split}
\end{align} 
where $u=a_s\sqrt{2\hbar m\omega_x\omega_y\omega_z/\pi}$, and the dimensionless relative position operator is represented with the bosonic operator $\hat{b}$ as $\hat{r}_x=(\hat{b}+\hat{b}^\dagger)/\sqrt{2}$. The matrix element of the interaction between harmonic basis $\ket{2m}$ and $\ket{2n}$ can be calculated as
\begin{align}
\begin{split}
 V_{mn} &= \int |\psi_{2m}(r)|^2 \sqrt{\pi}u\delta(r) |\psi_{2n}(r)|^2dr \\
 &= u\left(-\frac{1}{2}\right)^{m+n}\sqrt{\begin{pmatrix} 2m \\ m \end{pmatrix}\begin{pmatrix} 2n \\ n \end{pmatrix}}
\end{split}
\end{align} 
where $\psi_{2n}(r)=\frac{\pi^{-1/4}}{\sqrt{2^{2n}(2n)!}}\exp(-r^2/2)H_{2n}(r)$ is the wavefunction of the harmonic oscillator state $\ket{2n}$, and $H_{n}(r)$ is the $n$-th Hermite polynomial. The Hamiltonian in the even-number harmonic state basis,
\begin{align}
\begin{split}
 \hat{H} &= \sum_{n=0}^\infty \hbar\omega_x \left(2n+\frac{1}{2}\right) \ket{2n}\bra{2n} + \sum_{m,n=0}^\infty V_{mn}\ket{2m}\bra{2n}, \\
\end{split}
\end{align} 
can be diagonalized into $\hat{H}=\sum_n \tilde{E}_{2n}\ket{\widetilde{2n}}\bra{\widetilde{2n}}$ with perturbative eigenenergies
\begin{align}
\begin{split}
 \tilde{E}_0/\hbar\omega_x &= \frac{1}{2} + u' - (\ln{2})u'^2 + \cdots = 0.5 + u' - 0.69u'^2 + \mathcal{O}(u'^3), \\
 \tilde{E}_2/\hbar\omega_x &= \frac{5}{2} + \frac{1}{2}u' + \left(\frac{1-2\ln{2}}{8}\right)u'^2 + \cdots = 2.5 + 0.5u' - 0.048u'^2 + \mathcal{O}(u'^3), \\
 \tilde{E}_4/\hbar\omega_x &= \frac{9}{2} + \frac{3}{8}u' + \left(\frac{21-36\ln{2}}{256}\right)u'^2 + \cdots = 4.5 + 0.375u' - 0.01544u'^2 + \mathcal{O}(u'^3), \\
\end{split}
\end{align}
and perturbative eigenstates
\begin{align}
\begin{split}
 \ket{\widetilde{2n}} &= \ket{2n} + \sum_{m\neq n}\frac{V_{mn}}{2(n-m)\hbar\omega_x}\ket{2m} + \mathcal{O}(u'^2), \\
\end{split}
\end{align}
where $u'\equiv u/\hbar\omega_x$. We thus have
\begin{align}
\begin{split}
 \bra{\tilde{0}}\hat{r}_x^2\ket{\tilde{0}} &= \frac{1}{2}+\frac{1}{2}u'+\mathcal{O}(u'^2), \\
 \bra{\tilde{0}}\hat{r}_x^2\ket{\tilde{2}} &= \frac{\sqrt{2}}{2}+\frac{5\sqrt{2}}{16}u'+\mathcal{O}(u'^2), \\
 \bra{\tilde{2}}\hat{r}_x^2\ket{\tilde{2}} &= \frac{5}{2}+\frac{1}{4}u'+\mathcal{O}(u'^2), \\
 \bra{\tilde{2}}\hat{r}_x^2\ket{\tilde{4}} &= \sqrt{3}+\frac{5\sqrt{3}}{32}u'+\mathcal{O}(u'^2), \\
\end{split}
\end{align}
and
\begin{align}
\begin{split}
 \bra{\tilde{0}}\hat{r}_x^4\ket{\tilde{0}} &= \frac{3}{4}+\frac{9}{8}u'+\mathcal{O}(u'^2), \\
 \bra{\tilde{0}}\hat{r}_x^4\ket{\tilde{2}} &= \frac{3\sqrt{2}}{2}+\frac{23\sqrt{2}}{16}u'+\mathcal{O}(u'^2), \\
 \bra{\tilde{2}}\hat{r}_x^4\ket{\tilde{2}} &= \frac{39}{4}+\frac{45}{16}u'+\mathcal{O}(u'^2), \\
 \bra{\tilde{2}}\hat{r}_x^4\ket{\tilde{4}} &= 7\sqrt{3}+\frac{161\sqrt{3}}{96}u'+\mathcal{O}(u'^2). \\
\end{split}
\end{align}
In the main text, we define qubit states to be $\ket{\uparrow}\equiv\ket{\tilde{0}}$, $\ket{\downarrow}\equiv\ket{\tilde{2}}$, and Pauli matrices $\hat{\sigma}_z=\ket{\uparrow}\bra{\uparrow}-\ket{\downarrow}\bra{\downarrow}$, $\hat{\sigma}_x=\ket{\uparrow}\bra{\downarrow}+\ket{\downarrow}\bra{\uparrow}$. We thus obtain
\begin{align}
\begin{split}
 \hat{r}_x^2 &= \left(\frac{3}{2}+\frac{3}{8}u'\right)-\left(1-\frac{1}{8}u' \right)\hat{\sigma}_z+\left(\frac{\sqrt{2}}{2}+\frac{5\sqrt{2}}{16}u' \right)\hat{\sigma}_x+\mathcal{O}(u'^2), \\
 \hat{r}_x^4 &= \left(\frac{21}{4}+\frac{63}{32}u'\right)-\left( \frac{9}{2}+\frac{27}{32}u' \right)\hat{\sigma}_z + \left( \frac{3\sqrt{2}}{2}+\frac{23\sqrt{2}}{16}u' \right)\hat{\sigma}_x+\mathcal{O}(u'^2). \\
\end{split}
\end{align}
From these results, one can determine the drive strength $\lambda$ to prevent leakage to non-qubit states $\{\ket{\tilde{4}}, \ket{\tilde{6}}, \cdots\}$ by setting the coupling terms ($\bra{\tilde{2}}\hat{r}_x^2\ket{\tilde{4}}$, $\bra{\tilde{2}}\hat{r}_x^4\ket{\tilde{4}}$, etc.) to be smaller than the anharmonicity $\tilde{E}_4-2\tilde{E}_2+\tilde{E}_0$. While the above expressions are computed using perturbation theory to provide approximate parameter values, in our numerical simulations we do not use such perturbative treatments; instead, we use the exact solutions given in Ref.~\cite{Busch1998}.

\section{Quantum Gate Generation}

The original Hamiltonian of two atoms in an arbitrary modulated potential, with $\{V_k(t)\}=\{0,2V_0,0,0,0\}+\lambda\{\tilde{V}_k(t)\}$ in the main text, is
\begin{align}
\begin{split} 
 \hat{H} =& ~\hbar\omega_x\hat{a}^\dagger\hat{a} - \frac{\hbar\tilde{\omega}}{2}\hat{\sigma}_z \\
 &+ \lambda\epsilon_x \tilde{V}_1(t)(\hat{a}+\hat{a}^\dagger) \\
 &+ \lambda\epsilon_x^2 \tilde{V}_2(t)\left(\frac{1}{2}(\hat{a}+\hat{a}^{\dagger})^2+\hat{r}_x^2\right) \\
 &+ \lambda\epsilon_x^3 \tilde{V}_3(t)\left( \frac{3}{2}(\hat{a}+\hat{a}^\dagger)\hat{r}_x^2+\frac{1}{4}(\hat{a}+\hat{a}^\dagger)^3 \right) \\
 &+ \lambda\epsilon_x^4 \tilde{V}_4(t)\left( \frac{3}{2}(\hat{a}+\hat{a}^\dagger)^2\hat{r}_x^2+\frac{1}{8}(\hat{a}+\hat{a}^{\dagger})^4+\frac{1}{2}\hat{r}_x^4 \right) \\
 &+ \mathcal{O}(\epsilon_x^5) .\\
\end{split}
\end{align}
In the rotating frame of $\hat{H}_0=\hbar\omega_x\hat{a}^\dagger\hat{a}-\frac{\hbar\tilde{\omega}}{2}\hat{\sigma}_z$, the operators become time-dependent: $\hat{a}\rightarrow\hat{a}e^{-i\omega_xt}$ and $\hat{r}_x^2\rightarrow 3/2-\hat{\sigma}_z+\sqrt{1/2}\left(\hat{\sigma}_x\cos\tilde{\omega}t+\hat{\sigma}_y\sin\tilde{\omega}t\right)+\mathcal{O}(u')$.

\begin{table} [htbp]
\renewcommand{\arraystretch}{1.95}
\centering
\begin{tabular}{|m{0.05\textwidth}<{\centering}||m{0.23\textwidth}<{\centering}|m{0.55\textwidth}<{\centering}|m{0.11\textwidth}<{\centering}|}
 \hline
 Gate & Unitary operator & Modulating terms & $(\alpha,\gamma,\xi)/\omega_xT$ \\ 
 \hline
 D & $\text{exp}\left(\alpha\hat{a}^\dagger-\alpha^*\hat{a}\right)$ & $\tilde{V}_1(t)=V_0\sin\big(\omega_xt-\theta\big)$ & $\frac{\lambda}{8\epsilon_x}e^{i\theta}$ \\ 
 \hline
 R & $\text{exp}\left(i\gamma(\hat{a}^\dagger\hat{a}+c_1\hat{\sigma}_z) \right)$ & $\tilde{V}_2(t)= V_0$ & $\frac{\lambda}{4}$ \\ 
 \hline
 QR & $\text{exp}\left(-i\gamma\hat{\sigma}_\phi/2 \right)$ & $\tilde{V}_2(t)= V_0 \cos\big(\tilde{\omega}t-\phi\big)$ & $\frac{\lambda}{4}c_2$ \\ 
 \hline
 S & $\text{exp}\left(\frac{1}{2}(\xi^*\hat{a}^2-\xi\hat{a}^{\dagger2})\right)$ & $\tilde{V}_2(t)=- V_0\sin\big(2\omega_xt-\theta\big)$ & $\frac{\lambda}{8}e^{i\theta}$ \\ 
 \hline
 CD & $\text{exp}\left((\alpha\hat{a}^\dagger-\alpha^*\hat{a})\otimes\hat{\sigma}_\phi\right)$ & $\tilde{V}_3(t)=V_0\big[\sin\big((\tilde{\omega}+\omega_x)t-\theta-\phi\big) - \sin\big( (\tilde{\omega}-\omega_x)t+\theta-\phi \big)\big]$ & $\frac{3\lambda\epsilon_x}{16}c_2e^{i\theta}$ \\ 
 \hline
 CR & $\text{exp}\left((i\gamma \hat{a}^\dagger\hat{a})\otimes \hat{\sigma}_\phi \right)$ & \makecell{$\tilde{V}_2(t)= V_0\epsilon_x^2\big(3+c_3/c_2 \big)\cos\big(\tilde{\omega}t-\phi\big)$ \\ $\tilde{V}_4(t)=-2 V_0\cos\big(\tilde{\omega}t-\phi\big)$} & $\frac{3\lambda\epsilon_x^2}{4}c_2$ \\ 
 \hline
 CS & $\text{exp}\left(\frac{1}{2}(\xi^*\hat{a}^2-\xi\hat{a}^{\dagger2})\otimes\hat{\sigma}_\phi\right)$ & $\tilde{V}_4(t)= V_0\big[-\sin\big((2\omega_x+\tilde{\omega})t-\theta-\phi\big)-\sin\big((2\omega_x-\tilde{\omega})t-\theta+\phi \big)\big]$ & $\frac{3\lambda\epsilon_x^2}{8}c_2e^{i\theta}$ \\ 
 \hline
\end{tabular}
\caption{ A list of native gates. The gates include Displacement (D), Rotation (R), Qubit Rotation (QR), Squeezing (S), Controlled Displacement (CD), Controlled Rotation (CR), and Controlled Squeezing (CS). The magnitude of gate parameters $(\alpha,\gamma,\xi)$ can be mainly set by choosing a free parameter $\lambda$ and drive duration $T$.
}
\renewcommand{\arraystretch}{1} \label{S_table1}
\end{table}

Table.~\ref{S_table1} is a detailed version of the table in Fig. 3 of the main text. Here, we include explicit expressions of the modulating terms and gate parameters that generate the quantum gates. These coefficients are: $c_1\equiv \left(\bra{\uparrow}\hat{r}_x^2\ket{\uparrow}-\bra{\downarrow}\hat{r}_x^2\ket{\downarrow}\right)/2=-1+u'/8+\mathcal{O}(u'^2)$, $c_2\equiv \bra{\uparrow}\hat{r}_x^2\ket{\downarrow}=\frac{\sqrt{2}}{2}+\frac{5\sqrt{2}}{16}u'+\mathcal{O}(u'^2)$, and $c_3\equiv \bra{\uparrow}\hat{r}_x^4\ket{\downarrow}=\frac{3\sqrt{2}}{2}+\frac{23\sqrt{2}}{16}u'+\mathcal{O}(u'^2)$.

\section*{Higher order correction in Qubit- and Oscillator-Rotation gates }

Since the ideal unitaries in Table.~\ref{S_table1} neglect all time-dependent terms in the rotating frame of $\hat{H}_0$, we now take into account higher-order corrections arising from these terms, and derive more accurate expressions of the resulting unitaries. Generally, we start from the lab frame Hamiltonian with the choices $\tilde{V}_{k\neq2}(t)=0$, namely
\begin{align}
\begin{split}
 \hat{H} &= \hbar\omega_{x,0}\hat{a}^\dagger\hat{a}-\frac{\hbar\tilde{\omega}_0}{2}\hat{\sigma}_z + \lambda\epsilon_x^2\tilde{V}_2(t)\left(\frac{1}{2}(\hat{a}+\hat{a}^\dagger)^2+\hat{r}_x^2\right),\\
\end{split}
\end{align}
where we now use the $0$ subscript on $\omega_{x,0}$ and $\tilde{\omega}_0$ to denote the bare oscillator frequencies associated with this trap. In the rotating frame of $\hat{H}_0=\hbar\omega_{x}\hat{a}^\dagger\hat{a}-\frac{\hbar\tilde{\omega}}{2}\hat{\sigma}_z$, the Hamiltonian in the interaction picture becomes
\begin{align}
\begin{split}
 \hat{H}_I &= \hbar\Delta\hat{a}^\dagger\hat{a}-\frac{\hbar\delta}{2}\hat{\sigma}_z \\ 
 &\quad + \lambda\epsilon_x^2\tilde{V}_2(t)\left(\frac{1}{2}(\hat{a}e^{-i\omega_{x}t}+\hat{a}^\dagger e^{i\omega_{x}t})^2+c_1\hat{\sigma}_z+c_2(\hat{\sigma}_-e^{-i\tilde{\omega}t}+\hat{\sigma}_+e^{i\tilde{\omega}t})+c_2'(\ket{\tilde{2}}\bra{\tilde{4}}e^{-i\tilde{\omega}'t}+\text{h.c.})+\cdots\right) \\
\end{split}
\end{align}
where $\Delta\equiv\omega_{x,0}-\omega_{x}$, $\delta\equiv\tilde{\omega}_0-\tilde{\omega}$, $\hat{\sigma}_-=\ket{\tilde{0}}\bra{\tilde{2}}$, $\hat{\sigma}_+=\ket{\tilde{2}}\bra{\tilde{0}}$, $\hat{r}_x^2=c_1\hat{\sigma}_z+c_2\hat{\sigma}_x+c_2'(\ket{\tilde{2}}\bra{\tilde{4}}+\ket{\tilde{4}}\bra{\tilde{2}})+\cdots$, and $c_2'=\sqrt{3}+\frac{5\sqrt{3}}{32}u'+\mathcal{O}(u'^2)$. We define $\tilde{\omega}'$ to be the frequency difference between $\ket{\tilde{2}}=\ket{\downarrow}$ and $\ket{\tilde{4}}$.

\begin{figure*}[t]
 \centering
 \includegraphics[width=0.99\textwidth]{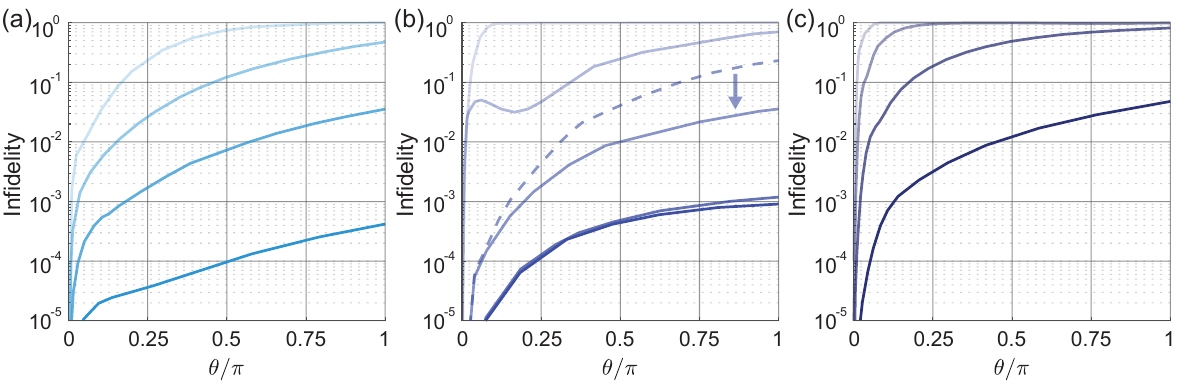}
 \caption{
 Infidelities of (a) Rotation, (b) Qubit Rotation, and (c) Controlled Rotation gates with higher-order corrections (solid lines). We use the initial state $\ket{\uparrow}\ket{\alpha}$, where $\ket{\alpha}$ is a displaced coherent state in the COM coordinate. From dark to bright (or bottom to top) lines, we use (a) $\alpha=3,6,9,12$, (b) $\alpha=0,3,6,9,12$, and (c) $\alpha=1.5,3,4.5,6$. The dashed line in panel (b) is the infidelity for $\alpha=6$, without accounting for higher-order corrections.
 }
 \label{S_fig1}
\end{figure*}

To obtain a more precise effective Hamiltonian, we perturbatively account for up to the second-order time-independent terms generated from rotating terms~\cite{James2007}. For the R gate, where $\tilde{V}_2(t)= V_0$ as shown in Table.~\ref{S_table1}, we obtain
\begin{align}
\begin{split}
 \hat{H}_\mathrm{eff}/\hbar &= \left( \Delta + \frac{\lambda\omega_x}{4} - \frac{\lambda^2\omega_x^2}{32\tilde{\omega}} \right) \hat{a}^\dagger\hat{a} + \left( -\frac{\delta}{2} + \frac{\lambda\omega_x}{4}c_1 - \frac{\lambda^2\omega_x^2c_2^2 }{16\tilde{\omega}} + \frac{\lambda^2\omega_x^2c_2'^2}{32\tilde{\omega}'} \right)\hat{\sigma}_z. \\ 
\end{split}
\end{align}
With appropriate choices of $\Delta$ and $\delta$ that cancel the undesired terms, one can obtain rotation gates generated by  $\hat{a}^\dagger\hat{a}$ or $\hat{\sigma}_z$. Similarly, for the QR gate, where $\tilde{V}_2(t)= V_0\cos(\tilde{\omega}t-\phi)$, we get
\begin{align}
\begin{split}
 \hat{H}_\mathrm{eff}/\hbar &= \left(\Delta-\frac{\lambda^2\omega_x^2}{64}\left(\frac{1}{2\omega_x-\tilde{\omega}}+\frac{1}{2\omega_x+\tilde{\omega}}\right)\right)\hat{a}^\dagger\hat{a} -\left(\frac{\delta}{2}+\frac{\lambda^2\omega_x^2c_2^2}{128\tilde{\omega}}+\frac{\lambda^2\omega_x^2c_2'^2}{128}\left(\frac{1}{\tilde{\omega}-\tilde{\omega}'}-\frac{1}{\tilde{\omega}+\tilde{\omega}'}\right)\right)\hat{\sigma}_z \\
 &\quad+ \frac{\lambda\omega_x}{8}c_2\left(\hat{\sigma}_-e^{-i\phi} + \hat{\sigma}_+e^{i\phi}\right), \\
\end{split}
\end{align}
with values of $\delta$ and $\Delta$ that eliminate $\hat{a}^\dagger\hat{a}$ and $\hat{\sigma}_z$ terms, we get qubit rotation gates $\hat{\sigma}_\phi=\hat{\sigma}_-e^{-i\phi}+\hat{\sigma}_+e^{i\phi}$.

In Fig.~\ref{S_fig1}, we show the simulated infidelities $1-\mathcal{F}$ of the rotation gates, analogous to the right panel of Fig. 3 of the main text for displacement and squeezing gates. We observe almost no effect from the higher-order corrections to the R gate, but generally find a significant reduction in the infidelity of the QR gate (see the dashed line in Fig.~\ref{S_fig1}(b)). Instead of a trivial initial state $\ket{\uparrow}\ket{0}$, we use displaced coherent states $\ket{\uparrow}\ket{\alpha}$ as our initial condition in Fig.~\ref{S_fig1}. We observe that the R and QR gates start to break down near $\alpha\sim 6$, owing to non-negligible higher-order terms. Additionally, the CR gate has various nonlinear terms that generate non-Gaussian dynamics on the quantum state, causing the fidelity drop even faster at $\alpha\gtrsim3$. All of these negative effects can be resolved by increasing $j_{max}$ for better control of higher-order terms $V_k(t)$, thus eliminating additional undesired terms in our time-averaged effective Hamiltonian.

\section*{Magnetic dipole-dipole interactions }

We first consider two magnetic dipoles aligned along the $z$ axis. The dipole–dipole interaction is given by
\begin{align} 
\begin{split}
    V_d(\vec{r}) &= V_{dd}\frac{1-3r_z^2/r^2}{r^3}, \\
\end{split}
\end{align}
where $V_{dd}=\mu_0/4\pi\times (g_F\mu_B)^2=2\pi\times 3.24\times 10^{-21} ~\text{Hz}\cdot m^3$, with Land\'e g-factor $g_F=-1/2$ for the internal state $^{87}$Rb $\ket{5^2S_{1/2},F=1,m_F=1}$, and Bohr magneton $\mu_B$. Here $r=|\vec r|=\sqrt{r_x^2+r_y^2+r_z^2}$ denotes the interatomic separation. 

To calculate the interaction strength for each qubit state, we start from the fact that the motional state associated with $\ket{\uparrow}$ ($\ket{\downarrow}$) can be well approximated by harmonic oscillator eigenstates for small $u$. For the relative motional wavefunctions, we take $\Psi_0(\vec r)=\psi_0(r_x)\psi_0(r_y)\psi_0(r_z)$ and $\Psi_2(\vec r)=\psi_2(r_x)\psi_0(r_y)\psi_0(r_z)$, where $\psi_n$ denotes the $n$-th harmonic oscillator eigenstate along each axis. The resulting interaction strength for each state is then given by
\begin{align} 
\begin{split}
    V_{d,\uparrow} &= V_{dd}\int d{\bf r} \frac{1-3r_z^2/r^2}{r^3}|\Psi_0({\vec{r}})|^2, \\
    V_{d,\downarrow} &= V_{dd}\int d{\bf r} \frac{1-3r_z^2/r^2}{r^3}|\Psi_2({\vec{r}})|^2. \\
\end{split} \label{eq:V00andV22}
\end{align}
The coupling between $\Psi_0(\vec{r})$ and $\Psi_2(\vec{r})$ is neglected as its effect averages out to zero in the rotating frame. To proceed, we first use the property
\begin{align} 
\begin{split}
    \int dx f(x)g(x) &= \int dk \tilde{f}(-k) \tilde{g}(k), \\
\end{split}
\end{align}
where $\tilde{f}(k)=\frac{1}{\sqrt{2\pi}}\int_{-\infty}^{\infty} f(x)e^{-ikx}dx = \mathcal{F}(f(x))$ is a Fourier transformed function. The expression $g({\vec{r_z}}) \equiv (1-3r_z^2/r^2)/r^3$ thus becomes
\begin{align} 
\begin{split}
    g({\vec{r}}) &= -\frac{\partial^2}{\partial r_z^2} \frac{1}{r} - \frac{4\pi}{3}\delta^{(3)}({\vec{r}}), \\
\end{split}
\end{align}
where the Dirac-delta function is introduced to resolve the singularity at $r=0$. Using the identity $\mathcal{F}[\partial_\alpha g(x)]=ik \mathcal{F}[g(x)]$, and
\begin{align} 
\begin{split}
    \frac{e^{-a r}}{4\pi r} &= \int \frac{1}{(2\pi)^3} \frac{e^{-i{\vec k}\cdot {\vec r}}}{k^2+a^2} d{\vec k} \\
\end{split}
\end{align}
for the Coulomb-type terms, we get
\begin{align} 
\begin{split}
    \tilde{g}({\vec k}) &= -\frac{\sqrt{2}}{3\sqrt{\pi}}\left(1 - 3\frac{k_z^2}{k^2}\right),  \\
\end{split}
\end{align}
where $\vec{k}=\{ k_x, k_y,k_z\}$ and $k=\sqrt{k_x^2+k_y^2+k_z^2}$. Additionally, the Fourier transforms of the harmonic wave functions are
\begin{align} 
\begin{split}
    \mathcal{F}(|\psi_0(x)|^2) &= \frac{1}{\sqrt{2\pi}}e^{-k^2x_0^2/4}, \\
    \mathcal{F}(|\psi_2(x)|^2) &= \frac{1}{\sqrt{2\pi}}\left(1 - k^2x_0^2+\frac{k^4x_0^4}{8}\right)e^{-k^2x_0^2/4}. \\
\end{split}
\end{align}
Eq.~\ref{eq:V00andV22} thus becomes
\begin{align} 
\begin{split}
    V_{d,\uparrow}/V_{dd} &= - \frac{4}{3\pi^2z_0^3} \int\int\int du_xdu_ydu_z \left(1 - 3\frac{u_z^2}{u^2}\right)\exp\Big(-(\alpha_x^2u_x^2+\alpha_y^2u_y^2+u_z^2) \Big), \\
    V_{d,\downarrow}/V_{dd} &= - \frac{4}{3\pi^2z_0^3} \int\int\int du_xdu_ydu_z \left(1 - 3\frac{u_z^2}{u^2}\right) \left(1-4\alpha_x^2u_x^2+2\alpha_x^4u_x^4\right) \exp\Big(-(\alpha_x^2u_x^2+\alpha_y^2u_y^2+u_z^2) \Big), \\
\end{split}
\end{align}
where $u_\eta\equiv k_\eta z_0/2$ for $\eta=x,y,z$, and $\alpha_\eta\equiv \eta_0/z_0$ for $\eta=x,y$. In the cylindrical coordinates with variables $u_r\equiv\sqrt{u_x^2+u_y^2}$, $\phi\equiv tan^{-1}(u_y/u_x)$, and $u_z$, we get
\begin{align} 
\begin{split}
    V_{d,\uparrow}/V_{dd} &= - \frac{4}{3\pi^2z_0^3} \int_0^\infty u_rdu_r \int_{-\infty}^{\infty} du_z\left(1 - 3\frac{u_z^2}{u_z^2+u_r^2}\right)e^{-u_z^2}\int_0^{2\pi}d\phi\exp\Big(-u_r^2(\alpha_x^2\cos^2\phi+\alpha_y^2\sin^2\phi) \Big), \\ 
    V_{d,\downarrow}/V_{dd} &= - \frac{4}{3\pi^2z_0^3} \int_0^\infty u_rdu_r \int_{-\infty}^{\infty} du_z\left(1 - 3\frac{u_z^2}{u_z^2+u_r^2}\right)e^{-u_z^2} \\
    &\quad\quad\quad\quad\quad\quad\quad \times \int_0^{2\pi}d\phi \left(1 - 4\alpha_x^2u_r^2\cos^2\phi + 2\alpha_x^4u_r^4\cos^4\phi\right)\exp\Big(-u_r^2(\alpha_x^2\cos^2\phi+\alpha_y^2\sin^2\phi) \Big). \\
\end{split}
\end{align}
The integrals over $\phi$ are
\begin{align} 
\begin{split}
    \int_0^{2\pi}d\phi\exp\Big(-u_r^2(\alpha_x^2\cos^2\phi+\alpha_y^2\sin^2\phi) \Big) &= 2\pi I_0 e^{-u_r^2\left( \frac{\alpha_x^2+\alpha_y^2}{2} \right)} , \\ 
    \int_0^{2\pi}d\phi \cos^4\phi\exp\Big(-u_r^2(\alpha_x^2\cos^2\phi+\alpha_y^2\sin^2\phi) \Big) &= 2\pi \left(\frac{3I_0 - 4I_1+I_2}{8}\right) e^{-u_r^2\left( \frac{\alpha_x^2+\alpha_y^2}{2} \right)}, \\ 
\end{split}
\end{align}
where $I_n\equiv I_n\left( u_r^2\left(\frac{\alpha_x^2-\alpha_y^2}{2}\right) \right)$ is the modified Bessel function of the first kind. The integral over $u_z$ is
\begin{align} 
\begin{split}
    \int_{-\infty}^\infty du_z \left( 1 - 3\frac{u_z^2}{u_z^2+u_r^2} \right) e^{-u_z^2} = -2\sqrt{\pi} + 3\pi u_r e^{u_r^2} \text{erfc}(u_r), \\ 
\end{split}
\end{align}
where $\text{erfc}(x)=1-\text{erf}(x)$ with $\text{erf}(x)$ the error function. In conclusion, we get
\begin{align} 
\begin{split}
    V_{d,\uparrow}/V_{dd} &= \frac{8}{3z_0^3} \int_0^\infty du_r u_r I_0 e^{-u_r^2\left(\frac{\alpha_x^2+\alpha_y^2}{2}\right)} \left( \frac{2}{\sqrt{\pi}} - 3 u_r e^{u_r^2} \text{erfc}(u_r) \right),  \\ 
    V_{d,\downarrow}/V_{dd} &= \frac{8}{3z_0^3} \int_0^\infty du_r \left( u_r I_0 - 4\alpha_x^2 u_r^3 \left(\frac{I_0 - I_1}{2}\right) +2\alpha_x^4 u_r^5 \left(\frac{3I_0 - 4I_1 + I_2}{8}\right)\right)  e^{-u_r^2\left(\frac{\alpha_x^2+\alpha_y^2}{2}\right)} \left( \frac{2}{\sqrt{\pi}} - 3 u_r e^{u_r^2} \text{erfc}(u_r) \right).  \\
\end{split}
\end{align}
With these expressions, we numerically compute dipolar interaction strengths in a 3D trap with trap frequencies $\{\omega_x,\omega_y,\omega_z\}=2\pi\times \{140,709,195\}$kHz, which gives us $V_{d,\uparrow}=-2\pi\times 62.8$Hz and $V_{d,\downarrow} = -2\pi \times 47.4$Hz, while the contact interaction strength is $2\pi\times 51.7$kHz. 

In the case where dipoles are aligned in the $y$-axis, one can basically use exactly the same calculation but just with switched $y$ and $z$. The interaction strength when dipoles are aligned in the $x$-axis can be automatically obtained since the interaction of the three cases should sum up to zero. In general, if the magnetic field is along $\hat{b}=\{b_x,b_y,b_z\}$ axis, with $b_x^2+b_y^2+b_z^2=1$, the difference between magnetic interaction strengths $\nu\equiv (V_{d,\uparrow} - V_{d,\downarrow})/\hbar$ is
\begin{align} 
\begin{split}
    \nu &=  2\pi \times \left[ b_x^2 (-104) + b_y^2( 119) + b_z^2 (-15) \right] \text{Hz}. \\
\end{split}
\end{align}
The maximum value of $\nu$ in the trap we used ($V_0=1$ mK) is thus $2\pi\times119$ Hz. For different trap depths, we obtain the maximum value as $\nu= 2\pi \times 119 ~\text{Hz}\times(V_0~/~1~\text{mK})^{3/4}$ when the magnetic field is along $y$-direction.

\section{Sensing dipolar interactions}

Here we describe details of the quantum sensing protocol shown in Fig. 4 of the main text. We obtain an analytical expression for the sensitivity. We start from the initial ground state $\ket{\uparrow}\ket{0}$, followed by QR gate $\exp(i\pi\hat{\sigma}_y/4)$, transforming the state into $\ket{+}\ket{0}$. The state then evolves under the interaction $\nu$ for a duration $t_d$, and becomes
\begin{align} 
\begin{split}
    \ket{\psi} &= \Big( \cos(\nu t/2)\ket{+} -  i\sin(\nu t/2)\ket{-} \Big)\ket{0} \\
    &= \Big( e^{-i\pi/4}\sin(\nu t/2 + \pi/4)\ket{+}_y + e^{i\pi/4}\cos(\nu t/2 + \pi/4)\ket{-}_y \Big) \ket{0}. \\
\end{split}
\end{align}
Next, we apply a squeezing gate (S), $\exp\left( \frac{\xi}{2}(\hat{a}^2-\hat{a}^{\dagger2}) \right)$, on the COM motional state, so the state becomes
\begin{equation}
\ket{\psi} = \Big( e^{-i\pi/4}\sin(\nu t/2 + \pi/4)\ket{+}_y + e^{i\pi/4}\cos(\nu t/2 + \pi/4)\ket{-}_y \Big) \ket{\xi} \\
\end{equation}
where $\ket{\xi}\equiv\exp\left( \frac{\xi}{2}(\hat{a}^2-\hat{a}^{\dagger2}) \right)\ket{0}$. Applying the CD gate $\exp\left( \alpha\hat{\sigma}_y\left( \hat{a}^\dagger-\hat{a} \right) \right)$ yields
\begin{equation}
\ket{\psi} = \Big( e^{-i\pi/4}\sin(\nu t/2 + \pi/4)\ket{+}_y \hat{D}(\alpha)\ket{\xi} + e^{i\pi/4}\cos(\nu t/2 + \pi/4)\ket{-}_y \hat{D}(-\alpha)\ket{\xi} \Big).  \\
\end{equation}
Unsqueezing this by applying $\exp\left( -\frac{\xi}{2}(\hat{a}^2-\hat{a}^{\dagger2}) \right)$, we get
\begin{equation}
\ket{\psi} = \Big( e^{-i\pi/4}\sin(2\pi\nu t/2 + \pi/4)\ket{+}_y \ket{\alpha e^{\xi}} + e^{i\pi/4}\cos(2\pi\nu t/2 + \pi/4)\ket{-}_y \ket{-\alpha e^\xi} \Big),  \\
\end{equation}
where $\ket{\pm \alpha e^{\xi}}\equiv \hat{D}(\pm \alpha e^{\xi})\ket{0}$.

Information on the COM quadrature $R_x$ (mean and variance) can then be measured via time-of-flight. These observables are given by
\begin{align} 
\begin{split}
\langle R_x\rangle &= \sqrt{2}\sin(\nu t) \alpha e^\xi ,  \\
\langle R_x^2 \rangle &= 2(\alpha e^\xi)^2 + \frac{1}{2}. \\
\end{split}
\end{align}
From these measurements, we can extract information about $\nu$ with a sensitivity given by: 
\begin{align} 
\begin{split}
    \Delta^2 \nu|_{\nu\to 0} &= \frac{1}{NM} \frac{ \langle R_x^2 \rangle - \langle R_x \rangle^2 }{|\partial \langle R_x \rangle / \partial \nu |^2}  \\
    &= \frac{1}{NMt_d^2} \left( 1 + \frac{1}{4\alpha^2e^{2\xi}} \right). \\
\end{split}
\end{align}
which is Eq.~(7) in the main text. 
Increasing the COM squeezing enhances the effective displacement amplitude, $\alpha e^{\xi}$, thereby suppressing the contribution of displacement-readout noise to the total uncertainty. As a result, the measurement becomes increasingly limited by the intrinsic standard quantum limit associated with the qubit phase accumulation.

To account for technical fluctuations in the trapping potential, we model the dynamics with an additional stochastic contribution to the Hamiltonian,
$
\frac{\hbar\omega_x}{2}\eta\left(\hat a^\dagger\hat a-\hat\sigma_z\right),
$
where the dimensionless noise parameter $\eta$ is drawn from a Gaussian distribution with zero mean and standard deviation $\sigma_\eta\simeq0.2\%$. Physically, these fluctuations induce shot-to-shot variations of the qubit and oscillator frequencies, leading to dephasing during the interrogation time $t_d$.
Including this effect, the sensitivity becomes
\begin{align} 
\begin{split}
    \Delta^2 \nu &\approx \frac{1}{NMt_d^2} \left( 1 + \frac{1}{4\alpha^2e^{2\xi}} \right)\times\exp\left( \sigma_\eta^2\omega_x^2t_d^2 \right). \\
\end{split}
\end{align}
The exponential factor captures the degradation arising from technical frequency fluctuations, which become increasingly important for larger trap frequencies and longer interrogation times. Consequently, the uncertainty increases in deeper traps, as illustrated in Fig.~4

\section*{Quantum Tomography of the motional state of two atoms}

It is known that full quantum tomography of a bosonic mode is available in a spin-boson system by utilizing a spin-dependent force. We first briefly introduce the protocol discussed in Ref.~\cite{Home2020}, where the Wigner function of the motional state of a single atom/ion is obtained.

Let's assume an atom with two relevant internal states $\{\ket{g},\ket{e}\}$ is initially prepared in the state $\ket{g}\ket{\psi}$, where $\ket{\psi}$ is the motional state we want to characterize. Obtaining the Wigner function $W(\gamma)$ for a complex number $\gamma$ is equivalent to obtaining the characteristic function, $\chi(\beta)\equiv\bra{\psi}\hat{D}(\beta)\ket{\psi}$ of the displacement operator $\hat{D}(\beta)\equiv\exp(\beta\hat{a}^\dagger-\beta^*\hat{a})$ where $\hat{a}$ is the bosonic mode annihilation operator, then obtain the Wigner function via $W(\gamma)=\frac{1}{\pi^2}\int\chi(\beta)e^{\gamma\beta^*-\gamma^*\beta}d^2\beta$. To measure $\chi(\beta)$, one can use the spin-phonon coupling (spin-dependent force) of the form $\hat{S}_x(\hat{a}+\hat{a}^\dagger)$, where $\hat{S}_x=(\ket{g}\bra{e}+\ket{e}\bra{g})/2$ is the Pauli-x operator. The protocol is as follows: first, a $\hat{S}_x$ rotation by angle $\theta$ is applied to the internal state, so that
\begin{align}
\begin{split}
 e^{-i {\hat S}_x \theta}\ket{g}\ket{\psi} 
 &= \frac{1}{\sqrt{2}}\left(e^{-i\theta/2}\ket{+}+e^{i\theta/2}\ket{-}\right)\otimes\ket{\psi},
\end{split}
\end{align}
where $\ket{\pm}\equiv \frac{1}{\sqrt{2}}\left(\ket{g}\pm\ket{e}\right)$. Next, the spin-dependent force term is applied for a displacement $\beta$, resulting in
\begin{align}
\begin{split}
 \hat{D}(\beta\hat{S}_x)e^{-i {\hat S}_x \theta}\ket{g}\ket{\psi} &= \frac{1}{\sqrt{2}}\left(e^{-i\theta/2}\ket{+}\otimes \hat{D}(\beta/2)\ket{\psi}+e^{i\theta/2}\ket{-}\otimes \hat{D}(-\beta/2)\ket{\psi}\right). \\
\end{split}
\end{align}
Lastly, $\hat{S}_z=(\ket{g}\bra{g}-\ket{e}\bra{e})/2$ is measured, which provides information of the characteristic function:
\begin{align}
\begin{split}
 2\langle\hat{S}_z\rangle &= \frac{1}{2}\bra{\psi} \Big(e^{i\theta}\hat{D}(-\beta)+e^{-i\theta}\hat{D}(\beta)\Big) \ket{\psi} \\
 &= \cos(\theta){\rm Re}[\chi(\beta)] + \sin(\theta){\rm Im}[\chi(\beta)],
\end{split}
\end{align}
and the real and imaginary parts of $\chi(\beta)$ can be obtained by varying $\theta$.

For two atoms with identical internal states $\ket{g}_1$ and $\ket{g}_2$, their motional state $\ket{\psi}$ can be measured by applying an analogous procedure. We define annihilation operators for the COM and relative coordinates as $\hat{a}_R\equiv(\hat{a}_1+\hat{a}_2)/\sqrt{2}$ and $\hat{a}_r\equiv(\hat{a}_1-\hat{a}_2)/\sqrt{2}$, respectively, so that the displacement operator in each coordinate becomes $\hat{D}_R(\sqrt{2}\beta)=\exp\left(\sqrt{2}(\beta\hat{a}_R^\dagger-\beta^*\hat{a}_R)\right)=D_1(\beta)D_2(\beta)$ and $D_r(\sqrt{2}\beta)=D_1(\beta)D_2(-\beta)$. 
Assuming the spin-dependent force is applied identically to each atom (i.e. $\beta \hat{S}_{x,1}(\hat{a}_1+\hat{a}_1^\dagger)+\beta \hat{S}_{x,2}(\hat{a}_2+\hat{a}_2^\dagger)$), and assuming the contact interaction is small and can be neglected during this measurement protocol, the state after the rotation and spin-dependent force becomes:
\begin{align}
\begin{split}
\prod_{j=1,2}\left(\hat{D}_j(2\beta\hat{S}_{x,j}) e^{-i {\hat S}_{x,j} \theta}\right)\ket{g}_1\ket{g}_2\ket{\psi} &= \frac{1}{2}\prod_{j=1,2}\left(e^{-i\theta/2}\ket{+}_j\otimes \hat{D}_j(\beta)+e^{i\theta/2}\ket{-}_j\otimes \hat{D}_j(-\beta)\right)\ket{\psi}\\
&= \frac{1}{2}\left( e^{-i\theta}\ket{+}_1\ket{+}_2\hat{D}_R(\beta\sqrt{2}) + e^{i\theta}\ket{-}_1\ket{-}_2\hat{D}_R(-\beta\sqrt{2}) \right)\ket{\psi} \\
& \qquad + \frac{1}{2}\left( \ket{+}_1\ket{-}_2\hat{D}_r(\beta\sqrt{2}) + \ket{-}_1\ket{+}_2\hat{D}_r(-\sqrt{2}\beta) \right)\ket{\psi} 
.
\end{split}
\end{align}
A joint measurement of both spin operators yields
\begin{align}
\begin{split}
 4\langle \hat{S}_{z,1}\hat{S}_{z,2} \rangle _\theta &= \frac{1}{2}\cos(2\theta){\rm Re}[\chi_R(\beta')] + \frac{1}{2}\sin(2\theta){\rm Im}[\chi_R(\beta')] + \frac{1}{2}{\rm Re}[\chi_r(\beta')]
\end{split}
\end{align}
where $\beta'=2\sqrt{2}\beta$, $\chi_R(\beta')=\bra{\psi}\hat{D}_R(\beta')\ket{\psi}$, and $\chi_r(\beta')=\bra{\psi}\hat{D}_r(\beta')\ket{\psi}$. Note that due to the indistinguishability of the two atoms, the relative coordinate can only occupy even-numbered states. Since $\hat{D}_r(\beta')-\hat{D}_r(-\beta')$ only includes odd numbered powers of $\hat{a}$ or $\hat{a}^\dagger$ terms, ${\rm Im}(\chi_r(\beta'))=\bra{\psi}\frac{\hat{D}_r(\beta')-\hat{D}_r(-\beta')}{2i}\ket{\psi}=0$, and thus ${\rm Re}[\chi_r(\beta')] =\chi_r(\beta')$. Therefore, we can obtain both $\chi_R(\beta')$ and $\chi_r(\beta')$ from the relations
\begin{align}
\begin{split}
 4\langle \hat{S}_{z,1}\hat{S}_{z,2} \rangle _\theta - 4\langle \hat{S}_{z,1}\hat{S}_{z,2} \rangle _{\theta+\pi/2} &= \cos(2\theta){\rm Re}[\chi_R(\beta')] + \sin(2\theta){\rm Im}[\chi_R(\beta')], \\
 4\langle \hat{S}_{z,1}\hat{S}_{z,2} \rangle _\theta + 4\langle \hat{S}_{z,1}\hat{S}_{z,2} \rangle _{\theta+\pi/2} &= \chi_r(\beta'). 
\end{split}
\end{align}
While these measurements enable tomography of either the COM or the relative motional states, they do not provide full quantum state tomography of the joint state. However, one can in principle learn more about the state by applying the above protocol in parallel with various control gates coupling the relative and COM motion, such as CD, CR, and CS gates.

\end{document}